\documentclass[journal]{IEEEtran}

\usepackage{mathptmx}       % selects Times Roman as basic font
\usepackage{helvet}         % selects Helvetica as sans-serif font
\usepackage{courier}        % selects Courier as typewriter font
\usepackage{MnSymbol}                       % not available on your system
\usepackage{grffile}
\usepackage{makeidx}         % allows index generation
\usepackage{graphicx}        % standard LaTeX graphics tool                          % when including figure files
\usepackage{multicol}        % used for the two-column index
\usepackage[bottom]{footmisc} % places footnotes at page bottom
\usepackage{subfigure}
\usepackage{multirow}
\usepackage{balance}
\usepackage{amsmath}
\usepackage{subfigure}
\usepackage{graphicx, subfigure}
\usepackage{authblk}
\usepackage{url}
\usepackage{array}
\newcolumntype{P}[1]{>{\centering\arraybackslash}p{#1}}
\usepackage{tikz}\definecolor{mediumturquoise}{rgb}{0.28, 0.82, 0.8}

\begin{document}

\title{Software-Defined VANETs:\\  Benefits, Challenges, and Future Directions}
\author{Wafa Ben Jaballah,~\IEEEmembership{Member,~IEEE,}, Mauro~Conti,~\IEEEmembership{Senior~Member,~IEEE,}  Chhagan~Lal~\IEEEmembership{Member,~IEEE,}
\thanks{Prof. Mauro~Conti, is with Department of Mathematics, University of Padua, Padua,~Italy.~e-mail:conti@math.unipd.it. The work of M. Conti was supported by
a Marie Curie Fellowship funded by the European Commission under the agreement PCIG11-GA-2012-321980. This work is also partially supported by the EU LOCARD Project under Grant H2020-SU-SEC-2018-832735. e-mail:conti@math.unipd.it}
\thanks{Dr. Wafa Ben Jaballah works at Thales, France. She is partially supported by the H2020 NRG-5 project under Grant H2020-ICT-2016-2-762013. e-mail:wafa.benjaballah@thalesgroup.com}
\thanks{Dr. Chhagan~Lal, is with Department of Mathematics, University of Padua, Padua,~Italy. He is also affiliated as Associate Professor in CSE Department at Manipal University Jaipur, Jaipur, 303007 India. e-mail:chhagan@math.unipd.it}}

\maketitle
\begin{abstract}
\boldmath
The evolving of Fifth Generation (5G) networks is becoming more readily available as a major driver of the growth of new applications and business models. Vehicular Ad hoc Networks (VANETs) and Software Defined Networking (SDN) represent the key enablers of 5G technology with  the development of next generation intelligent vehicular networks and applications. 
In recent years, researchers have focused on the integration of SDN and VANET, and look at different topics related to the architecture, the benefits of software-defined VANET services and the new functionalities to adapt them. However, security and robustness of the complete architecture is still questionable and have been largely negleted. Moreover, the deployment and integration of novel entities and several architectural components drive new security threats and vulnerabilities.
\par In this paper, first we survey the state-of-the-art SDN based Vehicular ad-hoc Network (SDVN) architectures for their networking infrastructure design, functionalities, benefits, and challenges. Then we discuss these SDVN architectures against major security threats that violate the key security services such as availability, confidentiality, authentication, and data integrity. We also propose  different countermeasures to these threats.  Finally, we discuss the lessons learned with the directions of future research work towards provisioning stringent security and privacy solutions in future SDVN architectures. To the best of our knowledge, this is the first comprehensive work that presents such a survey and analysis on SDVNs in the era of future generation networks (e.g., 5G, and Information centric networking) and applications (e.g., intelligent transportation system, and IoT-enabled advertising in VANETs). 

\end{abstract}


\IEEEpeerreviewmaketitle
\section{Introduction}
% By decoupling the control and data planes in VANETs, network intelligence and state can be logically centralized and the network infrastructure is abstracted from the applications. 
With the pervasive use of smart devices and advances in the development of wireless access technologies (e.g., DSRC, WiFi, 4G/LTE, and 5G), the Vehicular Ad hoc NETworks (VANETs) have become an accessible technology for improving road safety and transportation efficiency~\cite{Toor2008}. Due to continuous advancements in VANET technologies, it is seen as a network that can provide various services like vehicular cloud computing, surveillance, Internet of Thing (IoT) based advertising~\cite{Hidayet2018}, safety traffic management, to name a few. Although heterogeneous future architectures have been extensively investigated, the salient features of VANET (e.g., varying node density, high mobility) makes it challenging to coordinate VANETs to efficiently provide services with diverse Quality of Service (QoS) requirements. Hence, programmable networking architectures are becoming key enablers for VANETS to support inter-operation among underlying heterogeneous networks, conduct resource allocation tasks, and effectively manage a vast number of mobile nodes (or users) with heterogeneous smart devices.

\par In recent years, Software-Defined Vehicular Networking (SDVN) architectures have been emerged as a promising technology to simplify network management and enable innovation through network programmability. Thus, it gains significant attention from academia and industry. The SDN technology enables the decoupling of control and data planes in SDVNs, which provides: (i) an abstraction for VANET applications to the underlying networking infrastructure, and (ii) a logically centralized networking intelligence and network state. 

\par The convergence of SDN with VANET is seen as an important direction that can address most of the VANET's current challenges. In particular, the use of SDN's prominent features such as up-to-date global topology enabled dynamic management of networking resources and efficient networking services to enhance the user experience. These SDN features can meet the advanced demands of VANETs which includes high throughput, high mobility, low communication latency, heterogeneity, scalability, to name a few. Initially, the SDN paradigm have mainly used to improve the services in data centers, cloud computing, and access networks. However, nowadays SDN and OpenFlow techniques are being integrated into various next-generation wireless networks such as the Internet of Things (IoT)~\cite{Beraa2017}, Information-Centric Networking (ICN), and 5G~\cite{Yousaf2017}. Authors in~\cite{Yapp2010} evaluated Stanford ONRC OpenRoads project, which shows that SDN could help users to flawlessly travel across different wireless infrastructures~\cite{KOBAYASHI2014} on a large scale testbed for live video streaming application. The results showed that SDN makes it simple to perform handoff between Access Points and WiMAX stations, and it can tricast the video streams over both the networks (i.e., WiMAX and WiFi). Also, the cloud-medium access control (MAC)~\cite{Vestin2013}, a software-defined architecture for enterprise WLANs offer virtual wireless access points which dramatically improves its management.

\par In SDVNs, the SDN controller creates and maintains the repository which includes network-wide knowledge (e.g., network resources status and network topology). To this end, the controller gathers the networking topology information from underlying heterogeneous networking devices, e.g., Road-Side-Unit Controller (RSUC), Road-Side-Unit (RSU), and Base Station (BS). By using this global information at SDN controller, the VANET and SDN applications that reside at application layer could implement and enforce different network policies/configurations on the data plane devices through SDN controller via North-Bound Interface (NBI). In particular, the SDN controller can implement these configurations to the data plane elements by accessing these devices through South-Bound Interface (SBI) to produce coordinated and optimal decisions for the vehicles. 

\par Although there are various research efforts to provide efficient and feasible SDVN architectures, however, exploiting the full potential of SDN technology for new VANET applications is still at its initial stages. Additionally, during the design of SDVNs, we believe that security should be considered as a key requirement and an equally pressing issue. In contrast, the full transformation of existing VANETs into SDVNs will remain uncertain as long as the security, scalability, and data communication reliability issues of SDN are not been fully addressed. It is because the use of virtual centralization of network logic control (or intelligence) and the rapidly increasing cyber attacks make the emerging SDVNs more vulnerable to threats than the current VANETs. Moreover, the new entities and structural components that are being used in the current SDVNs are opening new attack surface and vulnerabilities which are unknown at the present stage. Consequently, it is required: (i) to perform a thorough investigation of the standardization efforts, and (ii) to address challenging issues (both, old and new) in the SDVNs.

\subsection{Contributions}
\label{contrib}

In this paper, first, we thoroughly discuss the working methodology of the state-of-the-art SDN-enabled Vehicular Network (SDVN) architectures and provide a generic design of vehicular network architecture integrated with SDN and other novel paradigms. We also investigate these architectures to identify their benefits and challenges against the traditional VANETs, mainly regarding the security and the communication reliability parameters. Then we present a set of potential requirements and primary enablers for a secure SDVN while we perform a security analysis of the existing SDVNs. Finally, an array of open research issues are presented that requires the attention of the researchers and professionals to establish a way forward towards a more secure and efficient SDVN that could enable the VANET usage in next-generation VANET applications. In particular, this paper provides the following key contributions.

\begin{itemize}
    \item We survey the state-of-the-art SDVN solutions for their benefits and challenges, mainly regarding security and performance of data communication processes. We believe that our survey will provide the required insights that will help to the possible development of a more secure and robust SDVN architecture. To the best of our knowledge, it is the first comprehensive survey that presents such a survey and analysis on SDVNs. Based on our study, we provide a generic design of  SDVN architecture.
    \item We present a detailed security analysis for an array of security threats along with their existing and possible countermeasures for the current SDVN architectures. Our report includes the security threats coming from the individual technologies (i.e., SDN only or VANET only), and the threats that result from the integration of SDN and VANETs (i.e., SDVNs). Finally, we discuss the lessons learned with the directions of future research work in this direction.

\end{itemize}

\subsection{Organization}

The rest of our paper is organized as follows. In Section~\ref{sec:backg}, we discuss the essential background overview about the VANET and SDN along with their benefits and challenges. In Section~\ref{sec:sdnvanet}, we provide the design of a generic SDVN architecture along with the survey of the state-of-the-art SDVN architectures. Additionally, the benefits and challenges of the existing SDVN architectures has been discussed in Section~\ref{sec:sdnvanet}. In Section~\ref{new_arch}, we present security analysis against an array of threats that could be launched on the SDVN architectures, and we discuss the existing and new possible solutions to countermeasure these threats. The lessons learned and the possible directions for future work are given in Section~\ref{diss_fut}. Finally, we conclude the paper in Section~\ref{concl}.

\section{Background Overview}
\label{sec:backg}
In this section, we provide a brief overview of VANET (in Section~\ref{intro-vanet}) and SDN (in Section~\ref{sdn}) technologies along with their working methodology, benefits, and challenges. Here, we only provide the details that are essential to understand the SDVN architectures that we have surveyed and investigated in the latter sections of this paper. The comprehensive overview of these two networking technologies is out of the scope of this paper, and we direct the interested readers to existing detailed surveys given in~\cite{Kreutz20153} and~\cite{Toor2008}.  

\subsection{Introduction to VANETs}
\label{intro-vanet}
VANET uses the Dedicated Short Range Communication (DSRC)~\cite{4539481} in order to provide wireless communication between vehicles ~\cite{4350110}. In this network, a vehicle can communicate directly with other vehicles forming vehicle to vehicle communication (V2V), or it can communicate with the road infrastructure  such as the road side unit (RSU) forming  a vehicle to infrastructure communication (V2I). The different communications between vehicles and infrastructures offer applications such as  accident prevention, traffic safety or traffic jams prevention\cite{4689254}. The main goal of sharing the information between vehicles is to provide safety messages in order to warn drivers about expected accidents, or to provide entertainment applications to passengers~\cite{BenJaballah14}. 
 Several challenges are facing researchers and developers when developing VANET applications, protocols, and simulation tools. Some researches have investigated the communication and networking aspects of VANET and addressed the security and privacy issues~\cite{BenJaballah14, LI20082803, BENJABALLAH20163, Benjaballah18CCNC}. Others focus on the routing protocols for VANET and their requirements to achieve better communication time with less consumption of network bandwidth~\cite{5247040, 7277098, SHAREF2014363}. Recently, some research works also investigate on providing more reliable and efficient services by integrating heterogeneous access networks such as LTE, 5G, NDN, Edge computing, and SDN~\cite{Gerla2014, ZHe2016, Liuu2017, Azizian2017}. 
\par
\begin{figure}[ht!]
\centering
  \includegraphics[scale = 0.65]{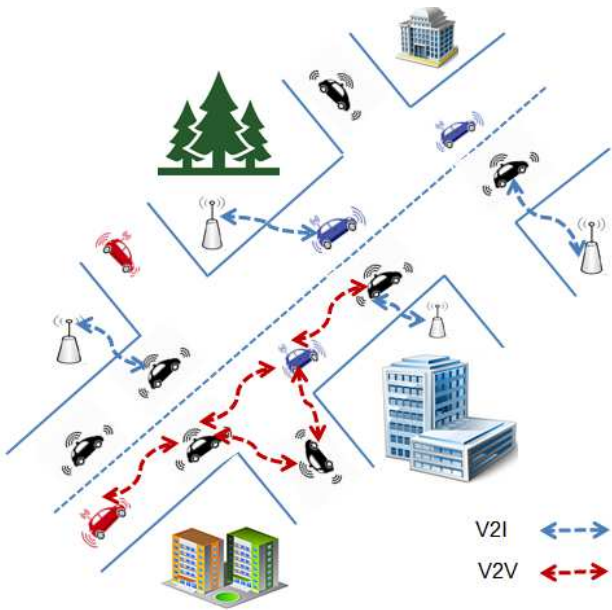}
\caption{\small{High level view of a generic Vanet architecture}}
  \label{Fig:vanetarch}
\end{figure}

 The Wireless Access in Vehicular Environments (WAVE) refers to the communication stack that is standardized by the IEEE. 
Vehicles communicate with each other as well as with RSUs through the WAVE wireless medium. A VANET architecture is mainly composed by three elements: the Application Unit (AU), the On-board Unit (OBU), and the RSU ~\cite{4350110}. In particular, the AU provides services, however the RSU hosts this application, and the OBU represents a device that uses the provided services.  Each vehicle in the system is equipped with an OBU to process the information, then it communicates a message with other vehicles or with RSUs ~\cite{4350110}.  
Three types of communications exist in VANET system (vehicles, RSU, the road infrastructure) that are: i) the intra-vehicle communication, ii) the V2V communication, and iii) the V2I communication. In the Intra-vehicle communication, an OBU communicates with different application units (AU) by providing a communication link. In the vehicle-to-vehicle communication, a vehicle communicates directly with another vehicle forming a one hop communication, otherwise if there is no direct connection, then vehicles execute a routing protocol to forward messages from one vehicle to another until it reaches the destination vehicle. The vehicle to infrastructure (V2I) communication consists of a vehicle that communicates with an RSU in order to process applications such as multimedia or video streaming ( please refer to Figure~\ref{Fig:vanetarch}).

\par Literature shows that a number of security threats exist in VANETs and a large number of these threats have been addressed~\cite{7286801}. In~\cite{Raya07}, the authors present three kinds of security threats in VANETs, including attacks on safety-related applications, attacks on payment-based applications, and attacks on privacy. They further proposed certain recommended mechanisms to resolve security issues in VANETs such as setting up tamper-proof hardware vehicles, and establishing public key infrastructure in the vehicular system. 
Moreover, in~\cite{BenJaballah14}, the authors propose some security requirements and an architecture for securing safety applications in VANETs. In~\cite{BENJABALLAH20163}, the authors focus on the position cheating attack. In particular, they determine the impact of different number of malicious vehicles on delaying the alert warning messages in vehicular communications. They identify the effective strategies and positions that could be used by adversaries in order to maximize the delay of the alert message.  In~\cite{Yan14}, the authors analyze security challenges and potential privacy threats specific to vehicular cloud, these includes the difficulty of establishing trust relationships and models among multiple actors that are due to intermittent short-range communications.

\subsection{Software Defined Networking}
\label{sdn}
The key concept of SDN~\cite{Xias2015} is decoupling of the \textit{control plane} and the \textit{data plane}. At control plane, a logically centralized entity called \text{controller} is used for monitoring and managing networking resources. The controller aims to improve the overall network performance (i.e., efficient communication and traffic control) by optimizing the usage of network resources. The data plane is a networking infrastructure, which is used for data forwarding, and it consists of forwarding devices and wired or wireless communication links. The SDN facilitates communication between devices from various vendors via standardized interfaces (e.g., OpenFlow). Thus, it provides ease in network monitoring and management, and it supports the design of programmable and flexible networking architecture.

\par The most commonly used programmable interface that provides communication between the entities of the two planes (i.e., control and data) is known as \textit{OpenFlow} protocol~\cite{McKeown2008}, and it runs on top of the Secure Sockets Layer (SSL). Apart from data and control planes, SDN also has a third plane called \textit{application plane}, which consists of third-party network services and applications. These SDN applications communicate with the SDN controller to express their essential requirements concerning security, QoS, or resource consumption, via an application-control interface. In particular, SDN uses the following two primary programming interfaces for inter-layer communications: (i) Control-Data Plane Interface called southbound API (e.g., OpenFlow Cisco OpFlex, and NETCONF), and (ii) Application-Control Plane Interface called northbound API (e.g., REST API).   

\par In a typical SDN network, each OpenFlow-enabled Switch (OF-Switch) connects to other OF-Switches and possibly to end-user devices that are the sources and destinations of traffic flows in the network. Each OF-Switch has multiple tables implemented in hardware or firmware that it uses to process (i.e., routing) the received packets. In particular, the controller modifies the content of the table called \textit{forwarding table}. Upon reception of a packet, the OF-Switch performs a lookup in its forwarding table to find the entry which specifies the corresponding action for the received packet. A table-miss occurs when there is no matching entry found for the packet, and it is processed as per the actions (e.g., send it to the controller through the southbound API or drop the packet) stated in the table-miss (or default) entry. The controller simply manages the network behavior by sending \textit{flowmod} packets that modifies the content of forwarding table at OF-Switches. The detailed discussion on SDN architecture and its technologies are out of the scope of this paper, for a comprehensive study on SDN, please refer to~\cite{Xias2015} and~\cite{Hayward2016}.

\subsubsection{Benefits and Challenges}
The SDN significantly simplifies network management by performing efficient resources usage with the help of the global network information, and it also ease the implementation of the networking services for SDN applications by abstracting the data plane from the applications and allowing them to enforce their dynamic requirements on data plane entities via logically centralized controller(s). Although SDN brings many benefits, the inherent characteristics (e.g., programmable SDN-based switches, the limited bandwidth of the southbound channel, and limited resources at SDN controllers) of SDN architecture also raises new security concerns. Below, we briefly summarize both the major \textit{benefits and challenges} in the usage of SDN technologies.

\begin{itemize}

    \item \textit{Support for heterogeneity and improved resource utilization:} With the use of its standard programmable interface such as OpenFlow, SDN architecture supports the device heterogeneity, i.e., networking devices coming from different vendors can interact with each other and with the control plane entities as long as they are configured with the open communication interfaces (e.g., OF protocol). The controller tries to always keep a current global view of the underlying networking infrastructure. Due to this, more than one real-world applications can share, through virtualization techniques, the same physical network to have a logically separate network. This makes the SDN re-usable as well as multi-purpose, i.e., it could be shared among different applications at the same or different point of times. In particular, the controller can instantiate multiple groups of logical OF-enabled switches on top of a single physical network in such a way that each physical entity could logically work for numerous applications, while each application will get a feel like the entity is working exclusively for it. Such instantiation of data plane entities pushes toward the maximum utilization of network resources by guaranteeing each application a customized performance which is based on their given requirements.
    
    \item \textit{Improved network security:} The controller can gather essential information about the network by communicating with the OF-Switches. These switches can collect the required information by performing network traffic analysis and using various anomaly-detection tools. Later, the controller analyzes and correlates the response from the data plane entities to create or update its global network view. Based on the analysis results, new configurations and policies to avoid the identified or predicted security threats can be installed in the whole network. Hence, these measures could help to improve the network performance and help in faster control and containment of identified security vulnerabilities.
    
    \item \textit{Single point of failures:} The centralized SDN controller, low bandwidth communication channel between the controller and OF-Switch and flow-table size limitation on OF-Switches makes the SDN vulnerable to an array of DDoS attacks. Moreover, the lack of (i) trust between data plane entities due to networks support for open programmability, and (ii) best practices particular to functions and components of SDN; remain major bottlenecks in the rapid and real-world adoption of SDN.  
    
    \item \textit{Slow propagation of bad information:} At OF-Switch, once a packet belonging to a certain traffic flow finds a match in the forwarding table, the switch \textit{knows} how to treat the remaining packets of the same flow. Therefore, it doesn't require any further interactions with the SDN controller. As it increases the traffic forwarding efficiently of the switch, but it also creates issues due to mobility, which makes the forwarding table rules inconsistent with the current network conditions. Therefore, the mismatch between the physical topology and global topology at controller causes packet losses (due to wrong forwarding information at OF-Switches) until the controller updates the forwarding table entries with new rules. 
\end{itemize}

\begin{figure*}[ht!]
\centering
  \includegraphics[scale = 0.33]{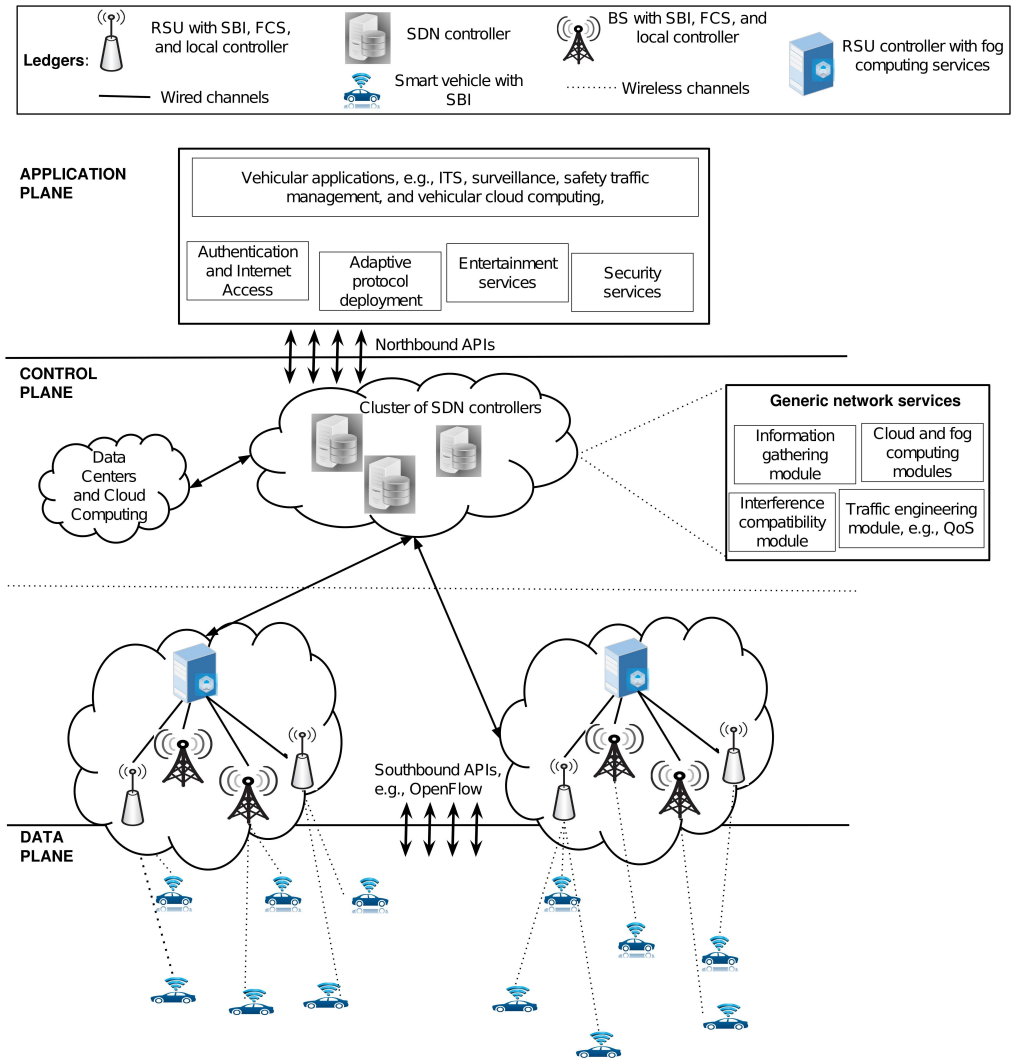}
\caption{\small{High level view of a generic SDVN architecture}}
  \label{Fig:sdvnarch}
\end{figure*}

\section{SDN based Vehicular Networks}
\label{sec:sdnvanet}

In this section, first we present a systematic survey of the state-of-the-art SDN-based VANET (SDVN) architectures. Then we will discuss the identified benefits and challenges concerning the performance gain and the robustness of data communication system in these architectures. In performance gain, we will provide an analysis of the impact on communication reliability, network throughput, network delay, resource utilization, and system complexity. While in robustness, we will discuss the impact concerning fault-tolerance, resistance to security attacks, and the size of attack surface.

\par Applying SDN in VANETs without any modification entails several challenges, mainly due to VANET's highly dynamic network topology which is attributed to the fast moving network nodes (i.e., vehicles). In particular, the key challenge in SDVN is to mitigate or minimize: (i) the huge management overhead on the controller, and (ii) the congestion on the control-data communication channel. With the frequently changing network topology, keeping the updated version of global network topology at controller is not only cost intensive and time consuming, but it also introduces inaccuracies in the received updated information. It is true even in the cases where the controller is able to keep an updated global topology view. However, some of the existing decentralized V2V and V2I communication protocols might not be able to use the benefits of SDN's global topology information system to its full extent. Hence, the network protocols might need redesigning to become adaptable with the SDN entities.

\subsection{Overview of state-of-the-art SDVN Architectures}
In this section, we provide a comprehensive survey on the existing works on SDVNs which includes a brief description of the proposed SDVN architectures along with their working methodology. We start our survey with the classification of the future SDVN architectures based on their integration with different paradigms (such as 5G, cloud computing, edge computing, and NDN), which has been done to improve their performance for specific and generic application domains. Table~\ref{tab-1} provides a brief summary of all the SDVN research works available in the state-of-the-art along with their key benefits and drawbacks. We broadly classify the surveyed SDVN architectures in following two categories. 

\begin{itemize}
    \item \textit{specific-purpose emerging vehicular networks based on SDN paradigm (SP-SDVNs)}: These architectures aim to improve specific VANET parameters such as network delay, QoS, access control, routing reliability, and security in data communication by using SDN technology. In some cases, to improve a set of parameters, the proposed architectures possibly use additional technologies (apart from SDN) such as 5G, named data networking (NDN), and fog or edge computing; and
    \item \textit{general-purpose emerging vehicular networks based on SDN (GP-SDVNs)}: These architectures aim to improve the overall performance of VANETs by integrating it with SDN and other technologies.
\end{itemize}

\begin{table*}
\caption {Review of State-of-the-art SDVN Architectures}
\centering
\scalebox{1.06}{
\begin{tabular}{|P{2.5cm}|P{1.7cm}|P{3.7cm}|P{3cm}|P{3cm}|}\hline
\textbf{SDVN Proposals and Architecture type} & \textbf{Additional Technologies Integrated} & \textbf{Description} & \textbf{Benefits} & \textbf{Drawbacks} \\ \hline
$[29]$, hierarchical & none  & routing protocol/Shortest travel time & rapid packet delivery, and low latency and overhead & maintaining global view at SDN controller \\ \hline 
$[19]$, centralized & none & SDVN to improve safety and surveillance services & communication efficiency & controller placement, security analysis \\ \hline 
$[20]$, centralized &  network slicing  &  heterogeneous vehicular communication & on-demand routing protocols and improved flexible and bandwidth utilization & benefits and challenges of slicing are not evaluated \\ \hline 
$[30]$, hierarchical & none  &  SDVN architecture to improve network flexibility and programmability & performance improvement is guaranteed even when connection is lost with main controller & insufficient performance evaluation\\ \hline 
$[31]$, hierarchical & fog computing & fog supported SDVN for autonomous driving, and automated overtake & improve delay sensitive and location awareness services  & effectiveness and correctness of the proposal remains unclear, fog, SDN, and VANET integration issues \\ \hline 
$[21]$, centralized & MEC, different access technologies  &  MEC-enabled SDVNs for reliable communication in Urban traffic management, and 4K streaming  & low latency and increase data rate   & connection loss with controllers and high mobility scenarios are not considered \\ \hline 
$[22]$, centralized & cloud computing  & SDVN for remote software updates in vehicles &  dealing with interference and hidden node issues & security, connection loss with controllers and high mobility scenarios are not considered \\ \hline 

$[32]$, centralized & artificial neural network & delay-minimization routing for SDVNs with mobility prediction &  Predicting mobility patterns in order to route vehicles & no security analysis, no solution when connection loss with controller occurs \\ \hline 
$[33]$, hierarchical & 5G & using SDN in 5G-enabled VANET to address DDoS attacks & provides a tradeoff between number of network services, dynamic topology, and network performance and security features & only weak security analysis \\ \hline

$[34]$, centralized & 5G & SDN-enabled buffer-aware multimedia streaming in 5G VANETs & better QoS during handover & scalibility and communication robustness\\ \hline

$[35]$, $[36]$, hierarchical & 5G, cloud/fog computing & fog-assisted SDN-based 5G VANETs to improve throughput and delay & reducing communication latency, improving scalability and flexibility & real-world performance evaluation\\ \hline

$[37]$, centralized & 5G & social-aware clustering protocol for SDN-based 5G-VANETs & reducing network congestion, improving packet delivery & security issues and connection loss with central controller are not considered\\ \hline
$[38]$, centralized & 5G & priority-based load balancer approach for data-offloading in SDVNs & improve scalability and traffic management & no evaluation is performed for mobility and security induced issues\\ \hline
$[39]$, centralized & 5G, MEC & improving V2V data offloading in 5G using SDN supported MEC architecture & contextual information based route discovery, V2V offloading & identifying accurate contextual information, vehicle privacy \\ \hline
$[40]$, centralized & none & advancements vehicular network technologies through SDN's unified network resource management approach & resource scheduling with a low cost communication overhead & no security analysis, no solution when connection loss with controller occurs\\ \hline
%$[41]$, hierarchical & & & traffic load balancing, conflict avoidance & \\ \hline
%$[42]$ & & & optimal share of network resources & \\ \hline
%$[43]$ & & & management of cooperative message dissemination  & \\ \hline
$[44]$, hierarchical & 5G, Cloud-RAN, fog computing & empowering real-time applications through SDN-enabled delay-sensitive, mobility-aware, and location-aware techniques &  management of cooperative message dissemination & integration issues with different technologies, evaluation in high mobility scenarios\\ \hline

\end{tabular}}
\label{tab-1}
\end{table*}

\par Before we start our survey on the state-of-the-art SDVN proposals, we give an overview of a generic SDVN architecture, which includes a comprehensive set of technologies and features that could satisfy a broad set of VANET applications. Most of the existing SDVN architecture designs are a subset of this generic SDVN design.  Figure~\ref{Fig:sdvnarch} provides a top-level view of our generic architecture for an SDN-based VANET along with its major components and their interactions. The data plane entities (e.g., smart vehicles) communicate with the control plane entities (global SDN and local RSU controllers, RSUs, and base stations) using the southbound APIs for coordinated and efficient communication. The controllers perform various functions such as routing, information gathering, and providing services to end-users based on the instructions and policies received through northbound APIs from the application layer entities. The controller provides an up-to-date network view to the application plane that helps it to manage various services (e.g., security, access control, mobility, and QoS) in the network. Specifically, vehicles communicate with their connected RSU or BS (i.e., RSUCs) to enable low latency local networking services and to maintain complete local knowledge. In parallel to serving the vehicle nodes, the RSUCs and BSs share the collected information about the vehicles and the transportation system to the global SDN controller. Based on the information received from the RSUCs, the network administrator that is sitting at the highest control of SDN controller builds a global view of the data plane entities (i.e., vehicles). To achieve VANET or SDN business services at the application layer of SDVN architecture, the applications interact with SDN controller (or RSUCs) through NBI protocols to program the data plane entities with the required and optimal network configurations.

\par To reduce the latency while providing time-sensitive services to the vehicles in SDVN, the local controllers are equipped with fog computing services (refer to Figure~\ref{Fig:sdvnarch}). These controllers perform the required processing for time-sensitive tasks on the received data before the data is sent to the cloud computing enabled data centers for further processing and analysis. Due to the availability of local controllers and Fog services (i..e, Fog enabled data plane elements) in SDVN, the architecture presented in Figure~\ref{Fig:sdvnarch} could operate in Hybrid Control Mode (HCM) which improves network performance and robustness. In HCM, the SDN controller takes partial control of the system, while sharing the remaining with RSUCs. This arrangement helps in situations where the local controller lost their connectivity with the global controller, and the vehicle nodes can still perform their networking tasks with the help of local controllers.

\par Researchers consider SDN as a promising paradigm to fill the gap between the heterogeneity caused by data plane entities (i.e., devices and wireless technologies) and the lack of route discovery schemes that could efficiently handle the dynamic topology changes in VANETs caused by the vehicle's inherent mobile nature. The dynamic topology causes packet losses in the network due to short lifetime links, therefore, routing protocols that could effectively analyze the link quality fluctuations are needed. To this end, authors in~\cite{Kalupahana2019} propose a new routing protocol for SDVN that takes into account the metrics which provides stable routes along with the lowest communication delay. Thus, it improves the packet delivery and decreases the latency. The proposed approach uses the availability of network-wide information at the SDN controller to identify the most stable routes with the shortest path between the communicating nodes. Authors in~\cite{Gerla2014} and~\cite{ZHe2016} proposed an architectural design and a set of services towards SDVNs. In~\cite{Gerla2014}, the proposed architecture aims to improve network flexibility and programmability, and it introduces additional network services, policies, and configurations for VANETs to cope up with the increasingly new requirements of advanced VANET applications. The authors demonstrate the feasibility and communication efficiency of SDVNs deployment by evaluating the performance of SDVN routing protocols with state-of-the-art MANET and VANET routing schemes. The goal in~\cite{ZHe2016} was to support rapid integration of innovative network services for efficient vehicular communications. The proposed architecture consists of different entities like vehicles, road-side units, and heterogeneous wireless technologies and devices, which are abstracted from the application layer through SDN controllers and SDN-enabled switches with unified interfaces. The authors also present some useful use cases to demonstrate how their architecture enables rapid network innovation. One advantage of the proposed architecture is that it allows programmability by selecting and deploying routing protocols on demand; another advantage is that it provides flexibility by using network slicing to isolate multiple tenants. To increase the robustness of the communication system in SDVNs, authors in~\cite{Correia2017} propose a hierarchical SDVN architecture intending to improve the performance in events where connection loss between vehicles and primary SDN controller are considered.

% [X8] [X14] [25]
%the below para contains the SDVNs plus edge computing plus 5G [X8] X[22] [X14] X[25]
\par To meet the requirements of future VANET scenarios which includes intelligent transportation systems (ITS), automated overtake and autonomous driving, the authors in~\cite{Truong2015} propose an SDVN architecture which uses fog computing services to provide services to delay-sensitive and location-aware applications. But, the paper lacks the validation of their proposed architecture. Taking a step further, authors in~\cite{Liuu2017} proposes a new SDVN architecture assisted by mobile edge computing (MEC) services to provide integration support for heterogeneous access technologies. The aim is to provide low-latency and high-reliability communication in the network. The validation of the proposed architecture concerning reliable data communication is carried out through a case study of \textit{urban traffic management}. The simulation results show that the architecture meets the application specific requirements concerning latency, reliability, and data rate.
Similarly, authors in~\cite{Azizian2017} propose an SDVN architecture with edge computing services to distribute software updates to vehicles in a flexible way. The architecture uses V2V beaconing information to create a global topology view at SDN controllers, which helps in systematic network management. Additionally, to tackle the challenges caused by interference and hidden nodes, the controller runs a technique that uses mathematical optimization models for assigning distinct operating frequencies to each vehicle. Finally, authors in~\cite{Tang2019} propose a delay minimization routing algorithm for SDVNs which uses machine learning techniques to predict mobility patterns in the network. Instead of relying on the delay minimization techniques such as edge services and continuously monitoring of vehicle location, the authors propose a framework that learns and anticipate the arrival rate of vehicles by employing the Artificial Neural Network (ANN) technology. The framework provides a statistical traffic model that gives traffic mobility estimates to the network managers, which enables them to perform effective routing decisions for vehicles.

%plus 5G [X11] [X13]
\par With the rapid advancements in next-generation technologies such as 5G and automotive applications, an integration between VANETs and 5G technology is envisioned by the network developers and service providers. To perform this integration efficiently, SDN is being used as a key enabler. For instance, authors in~\cite{Hussei2017} propose an integrated architecture of these three technologies (i.e., VANETs, SDN, and 5G) for providing a security-by-design approach in VANETs, and also to strike a fair balance between networking services, vehicle mobility, network performance, and security features. The proposed architecture is evaluated against an array of security threats (e.g., Denial of Service (DoS), resource exhaustion, and link layer discovery attacks) targeting either the SDN controllers or the vehicles. Additionally, the authors discuss several possible techniques to identify the source of attacks for their mitigation. Similarly, authors in~\cite{Laie2017} propose an SDN-enabled integrated VANET and 5G network architecture, which uses a novel buffer-aware streaming technique for real-time multimedia streaming applications. The proposal also aims to keep minimum communication latency and ensures good QoS during connection handovers between consecutive eNodeBs. To achieve adequate QoS, the SDN gathers information regarding user mobility and status of the player buffer, and the strength of the network signal is used to provide an efficient transmission strategy for multimedia streaming. 

%  [X12][X41] [44]
\par To further minimize the communication latency and to improve the QoS, authors in~\cite{Xge2017} and~\cite{Soua2018} propose a 5G SDVN architecture that also integrates the cloud and fog computing services to improve the network performance further. The proposal uses SDN to improve scalability and flexibility of vehicular networks, while fog cells have been introduced and fog computing is performed at the network-edge to lower the communication latency. The overall architecture is composed of various elements that include cloud-fog computing services, SDN and RSU controllers, RSUs, BSs, and vehicles. The controllers gather and share the state information of fog computing clusters to the cloud computing data centers. The data plane consists of network entities (e.g., vehicles, BSs, and RSUs), while the control plane consists of controllers (including RSU centers). The RSU center acts as a controller for an individual fog cell, which consists of vehicles and an RSU to avoid the frequent handovers between vehicles and RSUs. Vehicles in a fog cell communicate using a multi-hop relay. Authors in~\cite{Qi2018} examines possible techniques for integration of clustering algorithms with VANET supported 5G networks to minimize, the spectrum resources usage and the network congestion, and to improve the packet delivery ratio. Due to the challenges of finding an effective clustering algorithm which should be adaptive to dynamic VANETs, authors consider the SDN paradigm. In particular, the authors propose a social-aware clustering algorithm supported by the SDN paradigm for 5G-VANET systems. The algorithm exploits the social patterns such as future routes of vehicles, which helps to develop a prediction model to improve the stability of the clusters. 

%  [X16] [X43] 
\par After proposing different designs for SDVN architectures, the research community has shifted its focus towards evaluating and improving the data communication process and other networking services (such as QoS, security, and privacy) in these SDVN architectures. To this end, various solutions have been proposed that either exploit the unique in-built features of SDN or develop new techniques to achieve the application-specific or general requirements in SDVNs. For instance, authors in~\cite{Aujla2017} propose SDN based solutions to improve the data offloading mechanism in vehicular networks. The mechanism comprises load balancing and priority managing services that reside at the SDN controller. Additionally, the data offloading approach is used to minimize the network congestion which leads to higher network scalability with lower cost. Similarly, authors in~\cite{Huang2018} propose a data offloading technique for V2V communications in a cellular network inside an SDN-based Mobile Edge Computing (MEC) architecture. The proposed offloading method uses each vehicle's context information (e.g., location, speed, direction, and IDs), and it takes a centralized management approach (e.g., SDN controller services) for estimation and notification of communication routes between vehicles that are currently communicating by using a 5G network.

%  [X10] [X5] 
\par For better resource scheduling, the authors in~\cite{Zhangd2016} use the SDN as a unified resource manager in vehicular communications. The proposed solution takes into account the networking resources from the data plane to perform centralized scheduling at the control plane. The SDVN allows the network managers to choose optimal network interfaces whenever an application wants to send data. The integration of SDN with heterogeneous vehicular communication ensures a low-cost communication overhead. In~\cite{Radha2017}, authors present a service-channel allocation scheme adapted to SDVN communications. In particular, an SDN controller keeps a holistic view of network states and available spectrum resources. Based on this view, the controller could decide which channel to use for a service or traffic type. A key benefit from the SDN-enabled vehicular communications is that it avoids conflicts and interference. Additionally, the architecture supports load-balancing between different service channels. 

%Use of SDN technologies in 5G-enabled VANET - works [X1] [X6] [40]
\par To optimize network management, an SDVN communication technique is proposed in~\cite{Bozkaya2015}, which enforces an optimal share of network resources between the contended entities. A static distribution of network resources to RSUs can be ineffective in situations when traffic density under an RSU's coverage range increases because it forces the RSU to accommodate additional data, flows that could result in degradation in QoE of end users. To address such situations, the authors propose a mechanism for the management of data flows and transmission power and implement it on the controller. After identifying unsatisfactory vehicles (i.e., their QoE decreases), the model adjusts their signal levels by reducing the interference with RSUs. The key idea is based on a data-flow management technique that distributes unsatisfactory vehicles to each RSU. SDN has been also used to manage cooperative message dissemination in vehicular communications in~\cite{Liu2016}. In particular, the RSU controls data dissemination over V2V and I2V channels. The centralized RSU directs scheduling decisions for the vehicles with a set of instructions that specifies the channel to which it should tune to transmit or receive the data packets. This approach enables cooperative dissemination by using the SDN paradigm. Recently, authors in~\cite{khan2018} provide a hierarchical VANET architecture supported with 5G technology that also integrates the SDN's centralization and flexibility features. The architecture also uses Cloud Radio Access Network (C-RAN) with a 5G paradigm for effective allocation of resources using the SDN global view. Moreover, the architecture incorporates a fog computing framework to minimize the number of handovers (between vehicles and RSUs) that occurs over a defined period by using the zones and clustering techniques at the network-edge.    

\par In all the above discussed SDVNs, the placement of controller and open-flow switches is application specific. For instance, the SDN controller can be installed at RSUs, base stations, data centers, or vehicles, while the open-flow switch functionalities are usually installed in vehicles. As far as we know, all the proposed SDVN architectures did not consider integrating security modules or analyzing any of the security issues of their proposed architectures, which is a serious concern due to the use of VANETs in life-sensitive applications.

\subsection {Benefits and Challenges}
\label{BeneChal}
In this section, we will discuss the significant benefits and research challenges that we have extracted from our survey on the SDVNs architectures that was presented in the previous section.

\textbf{Benefits:} The benefits of SDN-enabled VANET are multiple, such as rapid network configuration, improving user experience by efficient resource utilization, minimizing service latency, and resistance to some inherited attacks of SDN or VANETs. Below, we discuss some of the major benefits of SDVNs.\newline
\begin{itemize}
    \item \textit{Optimized resource utilization -} The availability of global topology view helps the SDN controller to manage the network resource efficiently in SDVNs. For instance, when multiple wireless interfaces or configurable radios are available then the controller can choose better coordination of channel/frequency. Similarly, due to awareness of network resources, the controller can effectively choose whether and when to change the signal power of wireless interfaces to change the transmission range of the vehicles (or network nodes). For instance, when an SDN controller discovers that the node density under a RSU coverage area is too sparse, it sends instructions to the existing nodes in that area to re-configure their transmission power to achieve higher packet delivery ratio. Similarly, the SDN controller can take specific requirements from individual applications running on the top, and it can implement an optimal configuration for the network devices to meet these requirements. In particular, having an up-to-date network topology and resources view at controller opens-up new opportunities to improve the network performance by optimally allocating the resources based on the current network conditions. For example, in~\cite{Liu20156}, the authors utilize the global network knowledge for enabling GeoBroadcast in SDVNs. Another example, which shows how SDN improves the network resources usage is demonstrated in work presented in~\cite{Maio20167}. In traditional VANETs, any warning message by a source node in an intelligent transport system application is first sent to the nearest RSU. The RSU forwards it to the control center, from where the message is transmitted to all the other RSUs that resides in the geographical area. Finally, these RSUs will broadcast the message in their coverage area. This process of disseminating the warning message causes huge overhead concerning network bandwidth and latency. In \cite{Maio20167}, the authors perform the above operation in an efficient way by using the SDN technology which is as follows: (i) the source RSU forwards the first warning message to the SDN controller, (ii) the controller configures the routes (via flow entries) to the destination RSUs, and (iii) until modified, the same routes will be followed by all the upcoming warning messages to disseminate. In this way, the controller reduces network control overhead, communication latency, and bandwidth usage. The efficient use of network resources in SDVNs can significantly improve network performance for many target scenarios including both static (e.g., road accidents) and dynamic (e.g., make way for an ambulance).
    
    \item \textit{Fast and flexible network configuration -} The separation of control and logic plane in SDVNs provide support to the rapid and flexible network configurations. It will help to meet the varying requirements of the applications and to adapt the changes in network topology caused by vehicles mobility. For instance, due to the shortest path routing approach or due to dominant video applications which occupy large bandwidth on the route, congestion has occurred on a few selected forwarding nodes. With the help of up-to-date network information at SDN controller, such situation could be easily detected along with the IDs of the congested nodes, and the controller could perform traffic rerouting process, to avoid the congested nodes which lead to improvements in network resource utilization and performance, and it also reduces congestion points in the network. 

    \item \textit{Heterogeneous network integration -} In SDVNs, the controller provides the abstraction between VANET applications and networking infrastructure, which enables support for the integration of heterogeneous networks (e.g., wired and wireless) and communication technologies (e.g., DSRC, WiFi, LTE, and 5G) that reside at the data plane. The use of communication protocol such as OpenFlow greatly simplifies the interactions between the data plane and control plane entities. For instance, irrespective of the vendor and hardware configuration, an OpenFlow-enabled data plane switch could communicate with the controller through the well-defined southbound APIs.
    
    \item \textit{Minimizing service latency -} The use of SDN enables the optimal implementation and management of fog computing services at network edge routers, which significantly reduces the service latency for delay sensitive applications. In particular, SDN's programming flexibility feature provides great support for implementing fog computing services at SDN-enabled edge devices. The use of SDN global topology information allows dynamic re-configuration in flow tables of routers provide the support for the implementation of adaptive networking services, which helps in minimizing service latency. Thus, it leads to the improved end-user experience. For example, let us  suppose a vehicle $X$ goes outside the coverage area of an RSU. However, $C$ could receive service messages from a neighbor vehicle $Y$, and $Y$ is within the coverage area of RSU. In such a scenario, an RSUC can assign additional resources to vehicle $Y$ to support its increased needs (i.e., providing support to vehicle $X$). Authors in~\cite{Khelifi2018SecurityAP} provides such services to the lost vehicles by exploiting the inherent features of Information-Centric Networking (ICN).
\end{itemize}

\textbf{Challenges:} The state-of-the-art SDVNs faces issues in their large scale deployment in real-world applications, it is due to the following challenges in these architectures. 

\begin{itemize}
    \item \textit{Dynamic network topology -} The high vehicle mobility causes rapid changes in SDVN topology and fluctuations in radio communication channels. The frequent topology changes also hinder the real-time collection of the networking knowledge that is required at the controller to maintain a current view of the data plane resources. The delayed or inaccurate global perspective leads the controller to experience delays in distributing commands to network elements. Therefore, to support the rapid adoption of SDN paradigm in VANETs, it is required to develop mechanisms that could handle high network mobility management issues in target SDVNs. To this end, there exist few techniques (such as use of fog computing and local controllers at network edges) that tries to minimize the effect of network mobility in VANETs, however these techniques are not in the advanced stages and thus, these cannot be ported directly (i.e., without any optimizations) in SDVNs. At present, the most effective solutions to handle the mobility caused issues in SDVNs could be the ones, in which a vehicle's future directions can be predicted based on a set of metrics (e.g., velocity, past driving patterns, and GPS location) by applying machine learning tools. However, a correct and valid implementation of such solutions is challenging due to the privacy concerns and high deployment complexity.
    
    \item \textit{Broader flow rule definitions and policies -} In SDN, the data plane switches maintain forwarding tables which mainly consists of the following three entries: (i) packet forwarding rules, (ii) one or more action corresponding each rule, and (iii) a set of counters associated with a data flow to keep track of the number of packets or bytes handled. However, the existing flow rules and policies that govern the data communication in the SDN network needs to be enhanced to handle the essential demands of the broad range of new VANET applications. For example, the SDN controllers could offload some of the tasks to the RSUs and BSs which act as local (or lower level) controllers by sending general flow rules or policies instead of specific rules associated with a data flow. Latter, these local controllers could provide or install data flow particular rules and policies depending on their local knowledge of the network. Similarly, the RSUs and BSs could process the collected networking information locally for making some of the decisions, and also sent the same information to cloud data centers and SDN controllers via a southbound interface for global, long-term usage~\cite{Truong2015}.

    \item \textit{Security and privacy considerations -} In SDVNs, the SDN controllers manage network resources and also control various networking services (e.g., security, traffic management, and QoS services), therefore it is imperative to protect the SDN controllers from different cyber attacks. The propagation of malicious information to the controller from adversaries can lead to severe accidents. For example, DoS attacks can be launched to paralyze the operations of controllers, or controllers can be compromised through inside attacks. Hence, the security of the controller becomes a priority as it is a centralized decision making entity in SDVNs. Additionally, the new security vulnerabilities that might occur due to the integration of the VANET and SDN or other technologies with SDVNs needs to be investigated before the deployment of such hybrid architectures.  
    
    \item \textit{Interworking gaps among heterogeneous networks -} The coexistence of heterogeneous V2X networks require efficient interworking mechanisms that allow efficient communication between these networks. Also, the existing SDVN architectures are lacking standardized Eastbound/Westbound APIs and Northbound APIs for vehicular applications.
    
    \item \textit{Misbehavior of elements from different technologies (e.g., cloud, 5G, and ICN) involved -} The use of various technologies and architectures in realizing the next generation VANET applications also increases its attack vector. It is because misbehaving or vulnerability in any one of the integrated technology might affect the operations of the whole VANET. For instance, we have discussed above that the use of SDN controller adds a new set of security vulnerabilities in the network. Similarly, the drawbacks in other integrating technologies (e.g., cloud, 5G, and ICN) can significantly increase the threats in the integrated network. In~\cite{AK2017}, the authors present general security vulnerabilities and attacks for an SDVN. The work discusses the security implications of SDVN architectures at each layer. The layered (i.e., application, control, and data planes) architecture of SDVN must be secured in a way that the security solutions address cross-layer threats because the security breaches pertaining to one layer could cause harm to other layers as the layers are heavily dependent to each other. 
  
\end{itemize}

\section{SDVNs Security Analysis \& Countermeasures}
\label{new_arch}

In this section, we discuss the weaknesses of the state-of-the-art SDVNs against major security attacks that violates security services such as availability, confidentiality, authentication, and data integrity. We also discuss the existing countermeasures and provide possible solutions to handle the identified vulnerabilities. 

\begin{table*}
\caption {Attacks with varying technologies}
\centering
\scalebox{1.06}{
 \begin{tabular}{|P{6cm}|P{1.5cm}|P{1.5cm}|P{1.5cm}|} 
 \hline
  & \textbf{SDN} & \textbf{VANET} & \textbf{SDVN}  \\ \hline
 Control Plane Resource Consumption & $\checkmark$  &  &  $\checkmark$\\ \hline
 Network topology poisoning & $\checkmark$ &   & $\checkmark$ \\\hline
 Distributed DoS & $\checkmark$ & $\checkmark$  & $\checkmark$ \\\hline
 %Data to Control Plane Saturation Attack & $\checkmark$ &  & $\checkmark$ \\ \hline
 Rule Conflicts violating existing security policies & $\checkmark$ &  & $\checkmark$ \\  \hline
 On-board tampering &   & $\checkmark$  & $\checkmark$ \\ \hline
Privacy violation & $\checkmark$ & $\checkmark$  & $\checkmark$ \\\hline 
Forgery & $\checkmark$ & $\checkmark$  & $\checkmark$  \\\hline
%In-transit traffic tampering &  &  &  \\\hline
Jamming &  & $\checkmark$ & $\checkmark$ \\\hline
Impersonation &  & $\checkmark$ &  $\checkmark$\\ \hline
Application-based attacks &  & $\checkmark$ &$\checkmark$  \\ \hline
Malware attack injection & $\checkmark$ & $\checkmark$ & $\checkmark$ \\ \hline
Sinkhole attack  & $\checkmark$ & $\checkmark$ & $\checkmark$  \\ \hline
Sybil attack  & $\checkmark$ & $\checkmark$ & $\checkmark$ \\ \hline
Replay Attack  & $\checkmark$ & $\checkmark$ & $\checkmark$ \\ \hline

\end{tabular}}
\label{table_attacks}
\end{table*}

\par In Table~\ref{table_attacks}, we present the main attacks that threaten SDN systems, VANET, and whether they could be persistent in SDVN environments. When an attack targets the software-defined networks, it mostly impacts also the SDVN architectures such as the control plane resource consumption, the network topology poisoning, and rule conflicts. Moreover, attacks that are tailored against vehicular systems are most of them persistent on the SDVN architectures such as the on-board tampering, the jamming, and the application-based attacks. Attacks as the replay attack, the sybil, the sinkhole, malware injection, privacy violation, forgery, distributed DoS are persistent in SDN, VANET, and SDVN but with different requirements and impact on each technology. 
    
\subsection{Control Plane Resource Consumption.} 

Most of the SDVN architectures proposed in the literature~\cite{Gerla2014, ZHe2016, Correia2017, Hussei2017, Laie2017, khan2018} have been designed without security in mind. In particular, they are vulnerable to control plane resource consumption which is a major weakness in SDN. This attack is triggered when there are many requests to the control plane from the data plane. 
As in SDVN, the control plane is composed by different RSU controllers that can enforce flow rules, and then enables to control the network efficiently. However, this control mode can cause serious problems in particular due to many requests sent to the control plane. In SDVN like architectures, the RSU controller in~\cite{Gerla2014, Correia2017} should support a maximum of requests than usual. For instance, in some situations network packets in some RSU should wait until the vehicle deletes old flow rules. Finally, the impact of this attack is that it consumes resources of the control plane through the number of flow rules that could be handled, and the data plane through the number of flow rule entries. 

\par Possible countermeasures to the control plane resource consumption in SDVN architectures can be an adoption of current solutions in SDN such as in~\cite{Wang15,Wei15, Ambrosin2017, Shang17}. In~\cite{Wang15}, the authors proposed to keep both control plane and data plane functional even when there is a data-to-control plane saturation attack. In particular, they adopt the packet migration concept. Also, they used the data plane cache concept in order to reduce fake packets by distinguishing them from normal ones.  In a typical SDVN architecture, the two modules can be added at the controller level. In~\cite{Ambrosin2017}, the authors proposed LineSwitch, a solution based on two concepts related to probability and blacklisting. The solution provides both resiliency against SYN flooding saturation attacks and protection from buffer saturation. In~\cite{Shang17}, FloodDefender is based on three techniques that are the  table-miss engineering, the packet filtering, and the flow rule management. The main goal of this solution is to reduce the bandwidth jamming, and save the memory space of switches.  However, adopting these solutions without any modification might not be beneficial for the SDVNs due to the characteristics (e.g., high mobility, dynamic network topology, and resource constraints) of SDVNs data plane elements which are very different from the data plane elements of the SDN.

\subsection{Network Topology Poisoning.} 
The topology information is mostly related to upper-layer applications such as the packet routing, network virtualization and optimization, and mobility tracking~\cite{Khan17, Skowyra18}. When a network topology poisoning attack happens, this will cause dangerous situations since all the dependent services and applications will be affected. For instance, the packet routing can be affected and this will incur a man-in-the-middle attack or black-hole routing path. Another scenario illustrates the impact of an attacker succeeding to hijack the location of a network server in order to phish its subscribers. 
In a smart parking application using the SDVN architectures in~\cite{Soua2018}, the attacker can hijack the location of a controller to phish its service subscribers. Moreover, an attacker can even create black-hole routes by injecting false links in the topology.  Architectures in~\cite{Gerla2014, ZHe2016, Correia2017, Aujla2017, Zhangd2016,Soua2018} are vulnerable to the network topology poisoning.
Two known solutions TopoGuard~\cite{Hong15} and SPHINX~\cite{sphinx} detect these topology poisoning attacks via packet monitoring. TopoGuard detects false network links based on behavioral profiling. In particular, authors of TopoGuard propose a topology update checker module in order to monitor the network topology and validate topology updates~\cite{Hong15}. In SPHINX, the authors propose an anomaly-detection approach based on verifying the inconsistencies in network states. In~\cite{Skowyra18}, the authors propose an extension to TopoGuard called TopoGuard+. The solution monitors the characteristic control plane message patterns, and then defend against out-of-band port amnesia attacks.

\subsection{Distributed Denial of Service Attacks.} 

SDVN architectures~\cite{Yan15-DOS} are vulnerable to Distributed Denial of Service (DDoS) attacks. Since SDVNs architectures are split into three main functional layers: infrastructure layer (vehicles, RSU), control layer (RSU controllers), and application layer, then potential DDoS attacks can be launched on any one or more of these three layers. 
\par As a countermeasure to DDoS, one might use the solution in~\cite{SOM}, where authors propose  a machine learning technique for DDoS detection. In particular, the flow statistics are collected from the switches or vehicle sensors and then trained. 
However, using such a solution at vehicles could be problematic due to the resource constraint nature of sensors when compared with the generic SDN switches. Another solution Floodguard~\cite{floodguard} consists on preventing DDoS attack by using packet migration and data plane cache. The packet migration technique aims to protect both the controllers and the switches, and the data plane cache technique stores table-miss packets and differentiates anomalous packets from normal ones.

\subsection{Rule conflicts.}%. (see FRESCO)
Rule conflicts could have dreadful attacks in OpenFlow applications. For instance, some rules could be dedicated to quarantine a server that are overridden by a load-balancing application that may determine that the targeted host is the least-loaded server~\cite{Shin_fresco:modular}. 
Such vulnerabilities can be exploited by an attacker in the SDVN architectures presented in~\cite{Aujla2017, Zhangd2016, Gerla2014, Liuu2017, Qi2018}. Possible countermeasures can be adopted to solve the rule conflicts in SDN based applications. 
For instance, FortNOX~\cite{fortnox} detects rule conflicts that violate existing security policies, and offers an authorization enforcement in the controller kernel. One may install FortNOX features in the RSU controller in SDVN based architectures. 

\subsection{Privacy.}
In SDVN based architectures, various user relation information has to be protected such as the license plate, the position, and the driver's name; however the authorities should be able to reveal their identities in case of an accident or a dispute~\cite{He15}.  Conditional privacy preserving mechanisms in vehicular communications can be adapted to vehicular software architectures. In~\cite{Lin07}, the authors propose GSIS, a solution that integrates the group based signatures and ID based signatures, and offers security and privacy preserving mechanisms between different OBUs, and between OBUs and RSUs. 
In~\cite{Zhang08}, the authors propose a location privacy preserving authentication scheme based on blind signature. 
The scheme guarantees the location anonymity to the public. Using the proposed scheme, the probability of tracing a vehicle’s route is small. However, in SDVNs, the lack of secure communication channel (i.e., southbound interface) between the control and data plane, and disclosure on network resources stored at SDN or RSU controllers could expose the VANET users to various privacy risks. 

\subsection{Forgery.}
This attack consists on forging and transmitting false warning messages, in order to contaminate large portions of roads~\cite{BenJaballah14, BENJABALLAH20163}. For instance, an attacker can broadcast a forged GPS signal in order to mislead vehicles to get wrong location information. 
 Examples of traditional countermeasures against this attack is to ensure secure localization. In~\cite{Abu-Ghazaleh2005}, the authors present the triangulation as a technique to  determine the position of a vehicle from three reference points. Using this technique, attackers cannot decrease the distance between two neighboring vehicles. 
 In~\cite{capkun06}, the authors propose the verifiable multilateration  in order to determine the position of a vehicle from a set of reference points whose positions are known in advance.  Autonomous position verification~\cite{Leinmuller2006} is a mechanism to detect the impact of falsified position information in particular for position-based routing protocols at VANETs. It is based on various concepts such as the  maximum density threshold, and position claim overhearing. In~\cite{BenJaballah14}, the authors propose a secure distributed location verification to detect vehicles cheating about their positions. The detection mechanism does not rely on additional hardware but only on collaborative neighbors.

\subsection{Tampering.}
A vehicle that acts as a relay can disrupt communications of other vehicles, thus leading to in-transit tampering. Hence, the vehicle could drop or modify or corrupt messages. 
The on-board tampering consists on leveraging the data plane level of SDVN architectures composed by different vehicles. In particular, an attacker may modify data, tampering with the on-board sensing.  In order to detect tampered data packets, approaches are based on anomaly detection behaviors. For instance, in~\cite{ZLi08}, the authors propose an autonomous watchodog formation to ensure that watchdog nodes monitor the behaviors of the relaying nodes. 

\subsection{Jamming.} 
In a jamming attack, an attacker can partition the network even without compromising cryptographic mechanisms~\cite{Raya07}. Due to the broadcast nature of wireless communication, an attacker can jam the network by using a powerful transmitter. This attack could lead to prevent the reception of sensed data in case of a smart parking application. Moreover, the RSU controller will not be able to guide vehicles. In the SDVN based architectures, this attack could be mitigated. The RSU gathers and monitors the quality of channels, and then forward the report to a controller. This later selects the list of bad channels and asks the RSU to forward this list to all  deployed sensors \cite{Radha2017}. 

\subsection{Impersonation.} 
In this type of attack, an attacker  can masquerade as a police in order to mislead other vehicles to slow down or change direction~\cite{Raya07}. An  attacker can spoof also safety messages or service advertisements, and then impersonates roadside unit controllers. Moreover, an attacker can influence the route of its neighbor vehicles by spreading incorrect information about road conditions. 
Different approaches have been proposed in order to detect impersonation attacks~\cite{Xiao2006DIWANS, Chen09DCS, Grover2011, JFrancois14} in vehicular systems. In~\cite{Xiao2006DIWANS}, the authors propose a distributed approach, where every vehicle 
can verify the claimed positions of its nearby vehicles in order to detect misbehaving vehicles.  The proposed approach is based on  statistic algorithms to enhance the accuracy of position verification. The detection of  cheating nodes is confirmed when observing the signal strength distribution of a suspect vehicle over a period of time. In~\cite{JFrancois14}, the authors propose to trace back the potential sources of anomaly in the network. In particular, they propose a method to  identify the different switches composing the network path of an anomaly in the software defined network. 
In SDVNs, the impersonation attacks could be detected at the time of topology discovery,  that is performed by the SDN controllers by using the link layer discovery protocols.

\subsection{Application-based attacks.} 
In the following, we consider two specific vehicular applications such as smart grid and platoon management. We present how SDVN can be efficient to detect some of the attacks on these applications.   
    
\begin{itemize}
\item In a smart grid application, SDVN architecture can be used where data plane includes Electric Vehicles (EVs) and Electrical Vehicle Supply Equipments (EVSEs). Attacks on smart grid includes the network flooding, the topology poisoning, and transmission jamming.  When an  electric vehicle needs to connect to  the EVSEs,  an information message is sent to the controller in order to track the current network topology and status of the network.  The controller can  install forwarding rules, and it is able to detect attacks implied by anomalous behaviors in the smart grid application.  Anomaly detection systems such as the ones in~\cite{Shin_fresco:modular} can be added to the controller in order to monitor network traffic and detect compromised data. 

\item In platooning vehicular applications, attacks such as changing lane, merging, accelerating or decelerating, redirecting traffic or changing direction can be performed. In SDVN architectures, the controller can install the appropriate rules related to the acceleration/deceleration, merging/splitting and changing lane taking into account inputs from  traffic conditions and  events in the roads. Then a controller can collect information on road status and anomalous vehicle behavior by using exchanged messages. In particular, mechanisms such as the ones deployed in~\cite{Lin07} can be efficient to detect a misbehavior in platooning SDVN applications. In order to ensure better network utilisation, the RSU controller in SDVN based applications has the role to instruct the platoon leader to set different parameters. These parameters include the scheduling policy of data messages, the acceleration, or the deceleration. Furthermore, an RSU controller can detect attacks such as jamming, replay, or attacks targeting the management protocols. These attacks can induce maneuvers such as splitting,  merging, or  lane changing.

\end{itemize}

\subsection{Malware Attack Injection.} 
In SDVN based architectures, an attacker can maliciously inject a software that replicates itself through the different controllers and switches/vehicles. Remote attacks through Bluetooth or cellular communications allow the attacker to take control of a vehicle. One of the vulnerability limitations resides on the lack of message authentication of the controller area network (CAN).  In~\cite{vecure}, the authors propose a framework for vehicular systems that employs a trust group structure in order to authenticate messages of the CAN bus.

\subsection{Routing based Attacks.} 
In the following, we address the sinkhole, the sybil and replay attacks. In a sinkhole attack, an RSU  can execute this attack in order to instruct a portion of vehicles to route all traffic to it. This malicious RSU  behaves as a malicious gateway.  In~\cite{SHAFIEI2014}, the authors propose a centralized approach in order to detect infected regions in the network using geo-statistical model. Moreover, the authors propose a distributed monitoring approach to explore neighbors in order to detect malicious nodes. 
A sybil attack consists on creating multiple fictitious identities of vehicles in order to create an illusion of traffic congestion in the road. Approaches such as~\cite{Yu13, Xiao06, Zhou11}  analyze the signal strength distribution in order to detect sybil attacks. In~\cite{Yu13}, the authors propose a statistical method that verifies where a vehicle comes from. This approach uses statistical analysis over a period in order to improve the detection accuracy. In~\cite{Zhou11}, the detection of sybil nodes is done through passive overhearing by fixed points in the road. In a replay attack, the attacker first sniffs a message, and then reuses it in order to access to a restricted network.  Approaches such as~\cite{Lin07} for message authentication and authorization might be used in this context.

\section{Discussion and Open Issues}
\label{diss_fut}

In this section, first we summarize our findings along with the lessons learned that are gathered from our review on the state-of-the-art efforts of SDVN and its integration with other technologies (e.g., fog computing, vehicular cloud, NDN, and 5G) for supporting emerging vehicular network applications and services. Then we present the possible research directions along with the future issues and challenges. Some of these challenges we have covered in Section~\ref{BeneChal}, however, this section briefly covers the rest of the challenges and open research directions.

\par As confirmed by the large number of research works that we have discussed in our survey, the industry and academia are pushing towards the design and configuration of new SDVN architectures. The rapid push in this direction is the result of the emerging and innovative applications (e.g., 5G, Automated Transport Systems, and Internet of Vehicles) of VANETs that have stringent requirements concerning robustness, flexibility, latency (i.e., time constraints for critical real-time decision making), security, and privacy. In the literature, the researchers envisions the efficient deployment of these applications by using the SDVNs coupled with other next generation technologies such as mobile edge/fog computing, Name Data Networking (NDN), and Network Function Virtualization (NFV).

\par Although various architectures have been proposed in the literature to improve the communication reliability and security in VANETs, but the comprehensive investigation to evaluate the effectiveness and correctness of these architectures remains an open issue. In particular, the new security and privacy vulnerabilities that arises due to the coupling of new technologies (such as SDN, NFV, and mobile edge computing) with the existing VANET should be carefully studied. For instance, researchers should not only report the benefits of using SDN to improve VANET architecture, but the new issues (e.g., service latency, mobility, and securing the SDN controller) that are inherent to SDN and now hinders the performance of the SDVNs should also be investigated and discussed. Moreover, it is important to look at how dynamic real-time change, rapid on-demand growth (scalability), and integration of service context will play a key role in enabling successful deployment and avoiding performance visibility gaps in SDVNs. In the recent future, the VANET architecture will constantly be evolving to satisfy the rapidly growing requirements of its new applications. Therefore, we now present few research directions that could be exploited in this direction.

\begin{itemize}

\item \textit{Security of 5G slicing for V2X Services:} SDN and automotive systems are key enablers for 5G systems. This will hinder applications characterized by single or multi-tenancy. 
The diverging 5G V2X services  span from a single automated vehicle in a smart city, to enhanced real-time navigation systems on board. In traditional networks, different services can be supported in the same architecture, and  can be built without elasticity in mind. Moreover, these services share the same resources and are processed by the same network elements.  
 The concept of network slices has emerged as a novel technology that isolates network functions and resources. As defined by ~\cite{Rost17, Zhang17},  a network slice represents a collection of 5G network functions and specific parameters that are combined together in order to provide a specific use case or a business model. These resources and functions are tailored to a market's need on a shared infrastructure. The network slicing is based on virtualization, where the Network Function Virtualization (NFV) paradigm is based on the fact that network functions are not tied to the hardware. Hence, these network functions can be deployed as virtual network functions, and they run on different platforms. The SDN controller configures the different VNF and physical networks functions in one slice. Due to the features of V2V or V2I, different network slices can be presented such as: 1) slice for autonomous driving , 2) slice for tele-operated driving; 3) slice for vehicular infotainment applications; and 4) slice for vehicle remote diagnostics. 
Security and privacy challenges could be raised in one slice (intra-slice) or inter-slice communication. One should ensure that one slice cannot consume other slice's resources. Also, sharing a physical platform might also lead to attacks such as the side-channel attacks and privacy leaks. Moreover, an adversary might obtain capabilities to launch attacks to slices and on-going slices for instance in order to modify the configuration of other customer's slice instance, compromising a network function, or even terminate a slice. Hence, this will expose the services and network to disclosure and removal. We identify here the need to investigate security requirements and security solutions for network V2X slicing. Moreover, we should mention that important efforts are still needed from researchers and industries to design a complete approach to enable secure slicing in 5G vertical domains such as the automotive systems. 

\item \textit{Secure Function Chaining in SDVNs:} In virtualized environments, vehicles will require to communicate with the infrastructure in order to provide services such as traffic management, collusion avoidance, online gaming, etc. Hence, this will require the deployment of specific VNFs tailored for the network (as self-healing virtual functions: self-organization, terminal self-discovery, and mobility management) and  virtual functions for the intra-vehicle domain such as the virtual On-Board Unit function. On the other hand, in order to monitor traffic and avoid security attacks, network operators need to specify also security functions such as the virtual intrusion detection system, virtual Firewall, virtual Intrusion Protection system, and virtual DDoS. Different tenants might have different security requirements for their flows by considering a set of security functions their flow should pass by. Hence, one should consider the placement and the ordering of these VNFs. In~\cite{Bari15, GSun18}, the authors consider different approaches to secure VNF placement with respective to their ordering and instantiation in the traffic. This placement of secure functions throughout the flow traffic of tenants should be dynamic in order to cope with the mobility of vehicles and the different services that they provide. In~\cite{SLal17}, the authors present the security threats regarding the deployment and implementation of virtual network functions. 

\item \textit{Mobile Edge Computing  Security:} The Mobile Edge Computing (MEC)~\cite{ROMAN18} is based on a virtualized platform, and  enables applications to run at the edge of the network. The environment of MEC is characterized by low latency, proximity, high bandwidth, and location awareness. MEC opens up services to consumers and to industries to deliver their mission-critical applications. The automotive system can benefit from the MEC in order for instance to extend the vehicular cloud into a distributed mobile  environment. The MEC architecture enables data and applications to be placed close to the vehicles.  In the safety vehicular applications, the MEC applications can receive  messages  from the applications in vehicle or road side units. Then, these applications analyze these messages and  propagate  alert warning messages about road conditions to  nearby vehicles. Using MEC applications, a data is received in few milliseconds, thus allowing the vehicle or driver to react immediately. The MEC could host different VNFs in order to allow secure and trusted communications of services between vehicles or vehicles to infrastructures. However, the MEC comes up with security challenges related to: (i) the secure service chaining of different VNFs hosted in the MEC, (ii) the certification of VNFs at the MEC, and (iii)the use of distributed machine learning algorithms for intrusion detection at MEC to reduce the bottleneck and energy consumption of vehicles/sensors. 

\item \textit{Information Centric Networking (ICN) based Solutions:} Originally, the ICN was envisioned to address the pressing needs (e.g., device mobility, network scalability, access to information, and distributed content production) of today's Internet. However, due to its unique advantages that suits the various requirements of different network architectures, including SDN~\cite{SIRACUSANO2018}, 5G, and VANETs~\cite{Kalogeiton2018}~\cite{Arsalan2019}, the use of ICN paradigm is envisioned in these architectures as well. To this end, there are several preliminary solutions (i.e., ICN enabled SDVNs) have been proposed by using a widely known ICN instance, namely Named-Data Networking (NDN). The communication model of NDN replaces the traditional host-centric paradigm to a new information-centric one. Due to the various benefits that ICN provides, researchers have investigated its usage for addressing different VANET challenges~\cite{Grewe2016}~\cite{Kolonko2017}~\cite{Hussain2018}. For instance, authors in~\cite{Kalogeiton2018} propose a V2I communication architecture that exploits deployed RSU infrastructure for content retrieval in NDN-VANETs. The authors show that the use of NDN could provide improvements in VANET concerning mobility management, resource consumption, and faster content retrieval. We believe that the use of NDN in SDVNs or SDN in NDN-VANETs has significant potential to improve the VANETs, however, these domains still remains highly under investigated and needs significant work to move forward for real-world deployments. In particular, the new issues and challenges that arises from the combination of these three technologies needs to be fully understood and adequate solutions for the identified problems should be envisioned~\cite{Kaloge2017}.   

\item \textit{Mobility Management:} Providing efficient mobility management in SDVNs is important to keep a consistent and accurate global topology view at SDN controller, which is needed to correctly enable various networking functionalities (e.g., routing, traffic management, security services, and network virtualization) in the network. Although, the SDN provides network control which is flexible and programmable but its applicability to mobile networks (such as VANETs and 5G) is still in its infancy. Therefore, new mobility management techniques such as proactive mobility management algorithm implementation, and hybrid control plane switches to whom controller can delegates partial load for mobility management are needed~\cite{Khann2018}. One way to minimize the mobility induced communication challenges is to develop efficient and accurate mobility prediction models~\cite{Tang2019}. In SDVNs, firstly the availability of the network-wide topology at the SDN controller could help to predict accurate mobility of vehicles through advanced machine learning algorithms (e.g., artificial neural network (ANN)). Secondly, these prediction results can be used by the RSUs and BS during high mobilty events, to estimate the precise Expected Transmission Count (ETX) probability and end-to-end delay of each vehicle's request. Another option is to use ICN paradigm which supports efficient data retrieval in high mobility scenarios. ICN handles mobility issues because it facilitates data retrieval that is independent to the physical location of the source or producer of the data, hence it could be a key enabler for future vehicular networks~\cite{Signorello2016}. However, the ICN architecture also presents a new set of security vulnerabilities such as router cache poisoning, Interest flooding, and privacy violation attacks, these threats need to be properly investigated before its use in SDVNs.

\end{itemize}

\section{Conclusions}
\label{concl}

In this paper, we thoroughly investigate the state-of-the-art SDN enabled vehicular network architectures for their positive and negative impacts mainly in terms of security and privacy. Based on the existing SDVN architecture, we analyze different security vulnerabilities and attacks. We propose an array of open security research issues that require attention of industries and researchers to establish a way forward for more secure and efficient SDVNs. Moreover, we discuss the applicability of the existing solutions and propose the possible countermeasures to handle these attacks. At this point, we can safely conclude that the research on SDVNs is just beginning and SDN can support VANET to achieve its objectives that are needed to use it for next generation intelligent VANET applications and services. However, there are many issues that needs to be addressed before its practical deployment. This paper opens the debate for secure slicing in V2X communications, secure mobile edge computing,  mobility management, and information centric networking. Through the future research directions that we have raised, this work acts as a catalyst to address emergent security and privacy issues of future SDVN architectures.

\bibliographystyle{IEEEtran}
\bibliography{VANET_SDN}

\end{document}